\documentclass{caosp}
\usepackage{graphicx}

\articleNo{123}
\pubyear{2008}
\volume{38}
\volnumber{1}
\firstpage{1}
\received{November 30, 2007}
\accepted{December 10, 2007}

\begin{document}

\htitle{Segregation of isotopes of heavy metals due to LID ...}
\hauthor{A.\,Sapar, A.\,Aret, R.\,Poolam\"ae and L.\,Sapar}
\title{Segregation of isotopes of heavy metals due to
light-induced drift: results and problems}

\author{
        A.\,Sapar
      \and
        A.\,Aret
      \and
        R.\,Poolam\"ae
      \and
        L.\,Sapar
       }

\institute{
           Tartu Observatory\\
           61602 T{\~oravere}, Estonia \email{sapar@aai.ee}
          }

\date{November 30, 2007}

\maketitle

\begin{abstract}
Atutov and Shalagin (1988) proposed light-induced drift (LID) as a physically
well understandable mechanism to explain the formation of isotopic anomalies
observed in CP stars. We generalized the theory of LID and applied it to
diffusion of heavy elements and their isotopes in quiescent atmospheres of CP
stars. Diffusional segregation of isotopes of chemical elements is described
by the equations of continuity and  diffusion velocity. Computations of the
evolutionary sequences for abundances of mercury isotopes in several model
atmospheres have been made using the Fortran 90 program SMART, composed by the
authors. Results confirm predominant role of LID in separation of isotopes.
\end{abstract}

\section{Introduction}
About two decades ago Atutov and Shalagin (1988), Nasyrov and Shalagin (1993)
proposed light-induced drift (LID) as an effective physical mechanism for
diffusional separation of isotopes of chemical elements in the atmospheres of
CP stars. Thereafter in former papers (Sapar, Aret, 1995; Aret, Sapar, 2002;
Sapar {\it et al.}, 2005) we investigated some general features of the LID
phenomenon in the atmospheres of CP stars, assuming that the initial abundance
of the studied chemical element and its isotopes is constant throughout the
atmosphere and that the isotope mixture corresponds to the solar (or
terrestrial) one. Starting from corresponding initial and boundary conditions
and using our computer code SMART (Sapar, Poolam\"ae, 2003; Sapar {\it et
al.}, 2007), we have studied the evolutionary abundance changes of mercury and
its isotopes due to gravity, radiative acceleration and LID.

LID appears due to asymmetry of radiative flux in spectral lines. It can be
described as acceleration $a_{LID}$ additional to usual radiative acceleration
$a_{rad}$. The expression for $a_{LID}$ is similar to the formula for
$a_{rad}$ but instead of Voigt function there is its derivative relative to
wavelength. Effectiveness of LID depends on probability that electron stays
on the upper  level until the next collision (Sapar {\it et al.}, P27, these
proceedings).

\section{Separation of isotopes of heavy metals due to LID}
Asymmetry of flux in overlapping isotope lines generates different
accelerations to  isotopes, yielding their segregation. Isotopic spectral line
splitting is similar in most spectral lines and thus the effect of LID is
cumulative. LID causes rising of isotope with red--shifted lines and sinking
of isotope with blue--shifted lines. For heavy elements the effect of field
shift (due to nuclear volume) dominates over the mass shift in the opposite
direction. Thus, spectral lines of their heavier isotopes are shifted to
longer wavelengths. For heavy metals LID generally causes subsequent sinking
of the lighter isotopes and rising of the heavier ones, leaving finally only
the heaviest isotope in the atmosphere and its equilibrium abundance is then
determined predominantly by the usual radiative acceleration. However,
hyperfine splitting of spectral lines of isotopes with odd number of nucleons
somewhat complicates the picture of diffusional segregation.

High--precision spectral data and high--resolution model computations are
needed to model the LID. We have found that resolution $R = 5~000~000$,
corresponding to Doppler shift 60 m\,s$^{-1}$, can be considered as sufficient
for the computations. Values of collision cross-sections for atomic particles
determine effectiveness of LID and thus they are  needed with the highest
possible precision. However the data are yet of low exactness.

The formulae  used for LID computations are given in Sapar {\it et al.} (P27,
these proceedings). Diffusion coefficient by Gonzalez {\it et al.} (1995) was
used. The LID efficiency has been computed assuming long--range Coulomb
interaction between ions, the hard core impact model for neutrals and its
extension  outside the Debye sphere for impacts of ions with neutrals. We
specified boundary conditions by using the Lagrange 4th order interpolation
polynomials for all model layers.

\section{Computational software and results}
The software used is the Fortran code SMART composed by us for modelling
stellar atmospheres and studying different physical processes in them. The
code\-name is acronym of Spectra and Model Atmospheres by Radiative Transfer.

Program SMART enables to compute the  plain-parallel and static model stellar
atmospheres and corresponding emergent spectra of O, B and A spectral classes
in the temperature interval  from 9~000 to about 50~000 K. The lowest value of
temperature is due to the circumstance that absorption only  by H$_2$ and
H$^-$  molecules is taken into account. Restrictions in modelling are that the
atmosphere is chemically homogeneous and holds LTE.

Capabilities of program SMART include: isotopes segregation; getting detailed
radiative flux in all layers of stellar atmosphere; iterative correction of
initial model; relaxational formation of NLTE in line spectra; accelerations
of clumps in stellar wind; computation of detailed spectral limb darkening and
hence the spectra of rotating stars and  non-irradiated eclipsing binaries.

SMART is a compact and simple software. Its former FORTRAN 77 code has
been essentially improved and refactored to Fortran 90. Using the code several
evolutionary segregation scenarios for mercury isotopes in quiescent
atmospheres of CP stars have been computed. Computation of one time step takes
approximately 15 min  on a PC with CPU  3.2 GHz, 2 GB RAM.

Model atmospheres have been computed with SMART, using sampling for
moderate spectral resolution ($R$=30 000). Smooth transition from spectral
line series to corresponding continua is a special feature of the code.  It
was achieved by introducing probability functions of continuum depression
which are complementary to corresponding existence probabilities of
high-excitation (Rydberg) electron states. Spectral line data from Kurucz file
\texttt{\small gfhyperall.dat} have been used in computations. Spectral line
data for Hg have been compiled   using different sources and improved by
adding isotopic splitting to all Hg lines. The line list used by us contains
about 700 resonance and low excitation spectral lines for HgI,  HgII and
HgIII, i.e. for ion species, which are  most important for LID.

Formation of evolutionary stratification of Hg isotopes has been computed
for a set of three effective temperatures ($T_{eff}=$ 9~500~K, 10~750~K,
12~000~K) and three initial Hg abundances ($\rho^0=$ solar, solar + 3dex,
solar + 5dex). We assumed homogeneous initial abundance of Hg throughout the
atmosphere and solar (terrestrial) ratios of isotope abundances. Possible
presence of  stellar wind and  microturbulence, both reducing or even
cancelling the diffusional segregation of isotopes, has been ignored. The
longest evolutionary sequence (500 time steps, \`a 1 year) has been computed
for model atmosphere with parameters $T_{eff}$=10~750~K, $ \log g=4$ and
initial Hg abundance  solar + 5dex. Due to much longer  free paths  in  higher
atmospheric layers the isotope segregation proceeds there essentially quicker
than in the deeper and denser ones. Drastically different diffusion time
scales in upper and lower atmospheric layers cause an essential computational
problem: time-steps have to be chosen small enough to ensure stability of
algorithms in upper layers, but a very large number of time-steps, which is
necessary therefore, generates only small changes in deep layers.

\begin{figure}[bth]
  \centering
\includegraphics[viewport=3 15 355 252, clip, width=.5\textwidth]{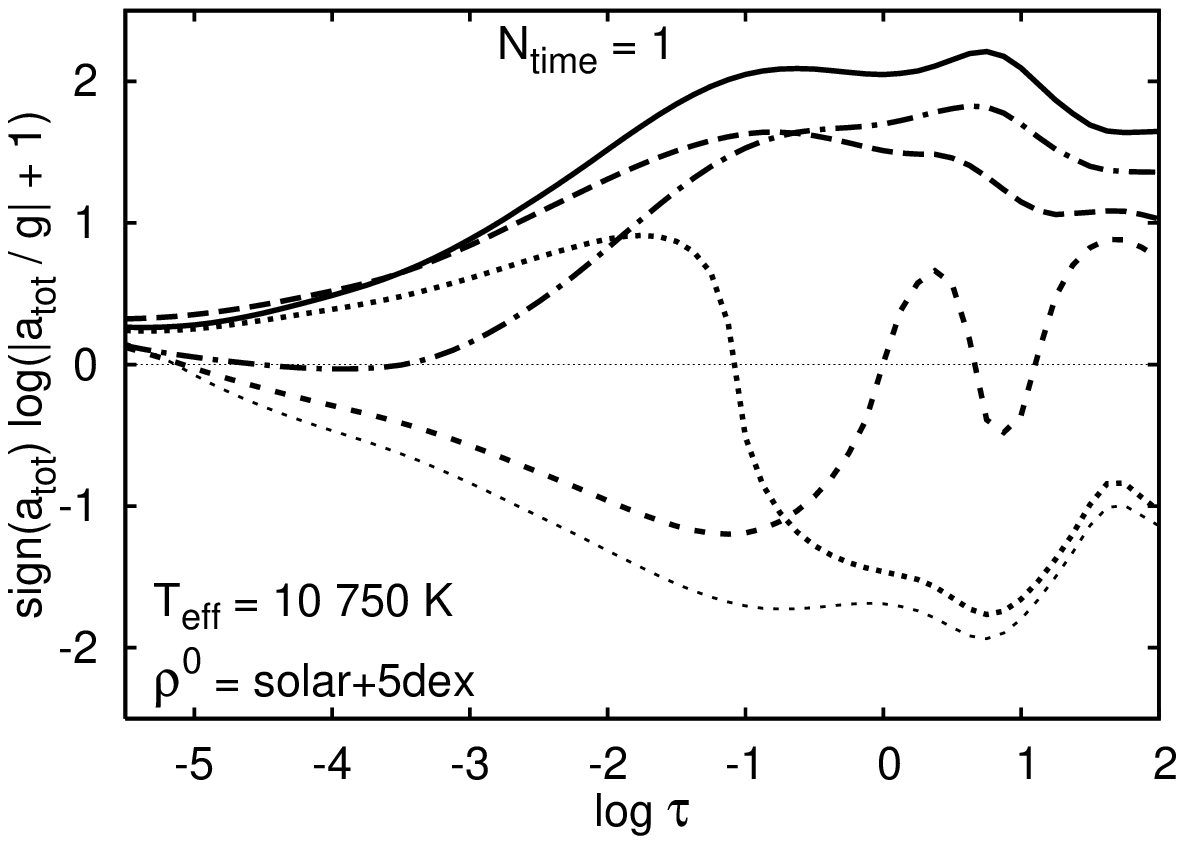}\includegraphics[viewport=3 15 355 252, clip, width=.5\textwidth]{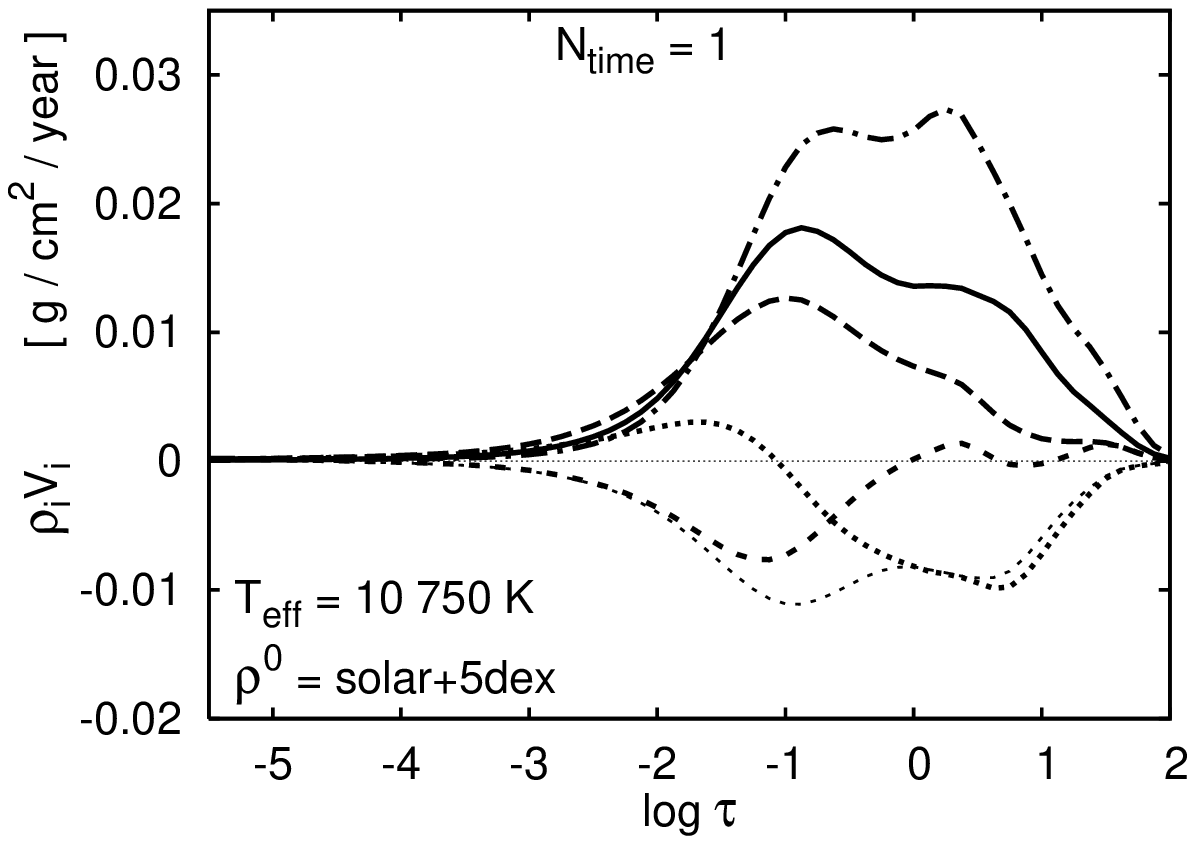} 
\includegraphics[viewport=3 -1 355 252, clip, width=.5\textwidth]{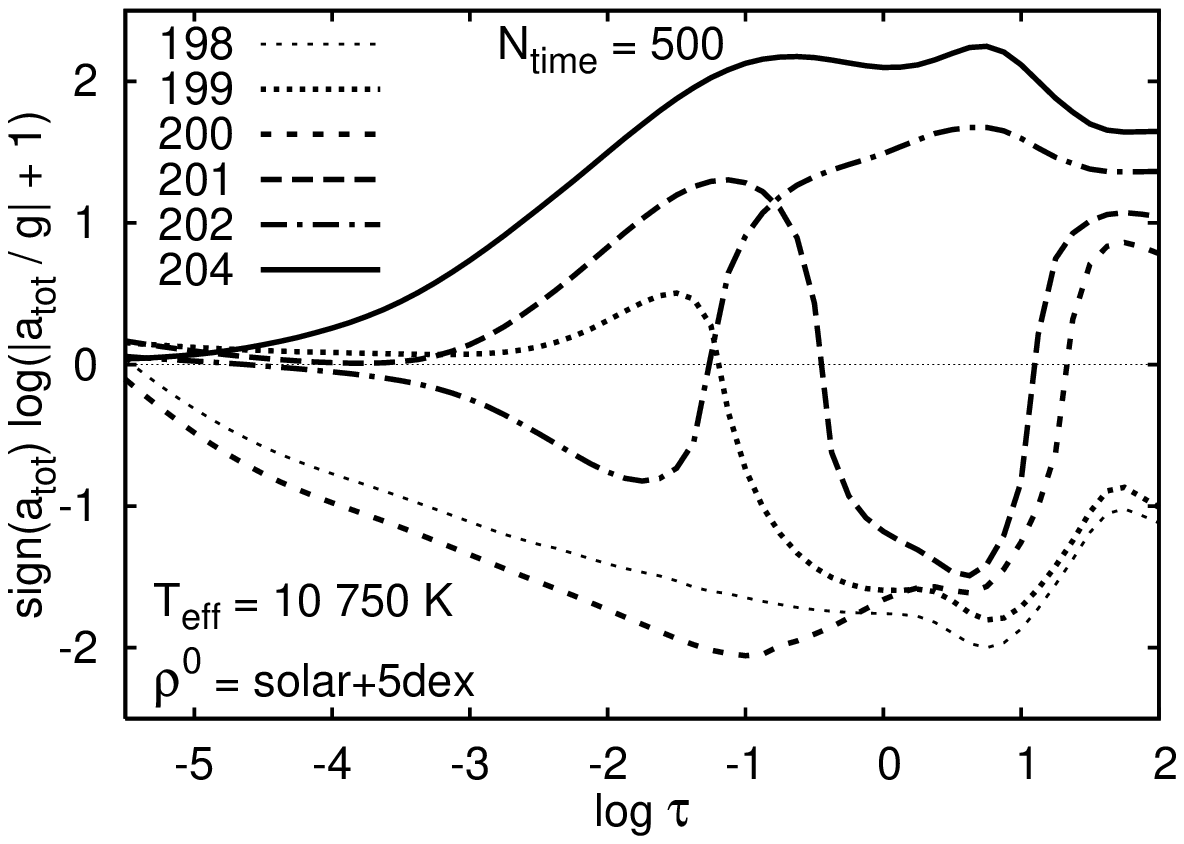}\includegraphics[viewport=3 -1 355 252, clip, width=.5\textwidth]{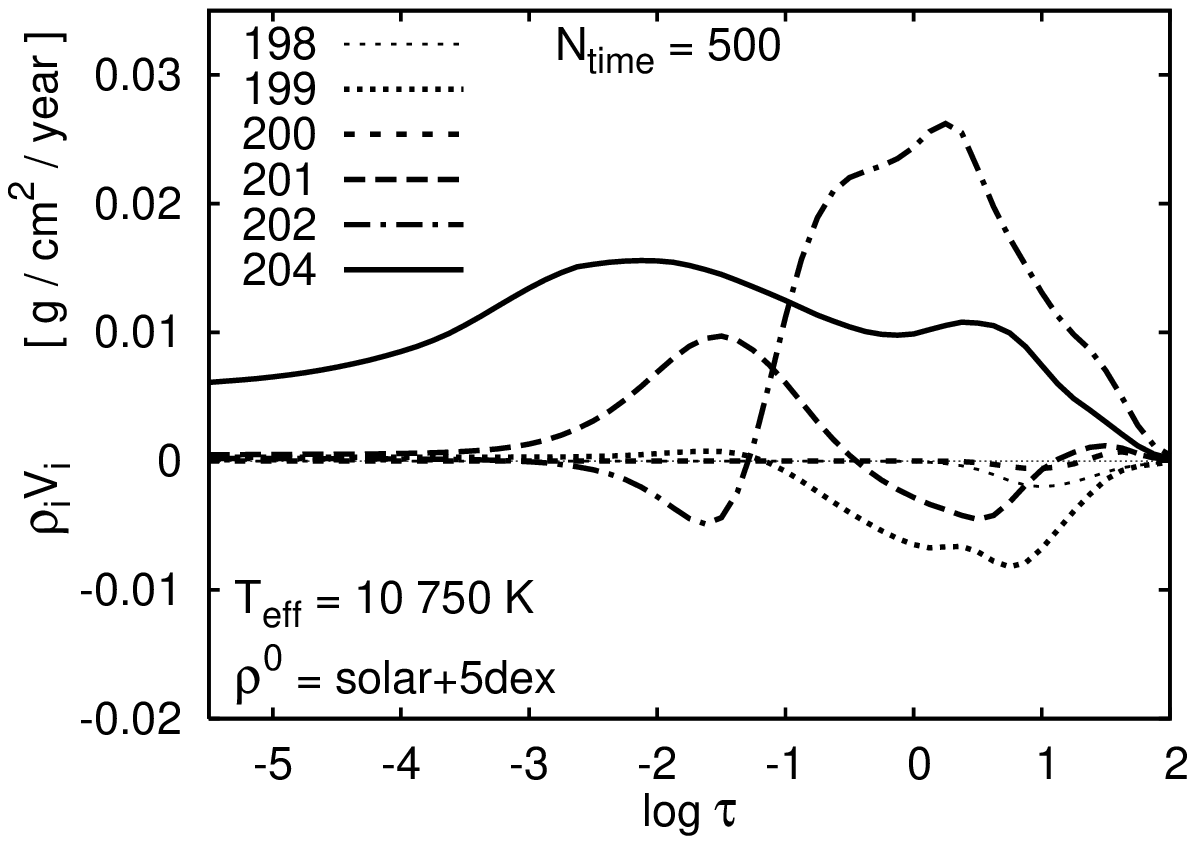}
  \caption{Change of acceleration $a_{tot} = a_{rad} + a_{LID}$ (left column)
   and of flows of Hg isotopes (right column) from the first to the 500th time-step.
   The ratio $a_{tot}/g$ is given in modified logarithmic scale.
   Note the complicated change of the curves for intermediary isotopes
   and  evolutionary damping of the flows.}
   \label{fig:C_sapar1}
\end{figure}

\begin{figure}[bth]
  \centering
\includegraphics[viewport=5 15 355 252, clip, width=.5\textwidth]{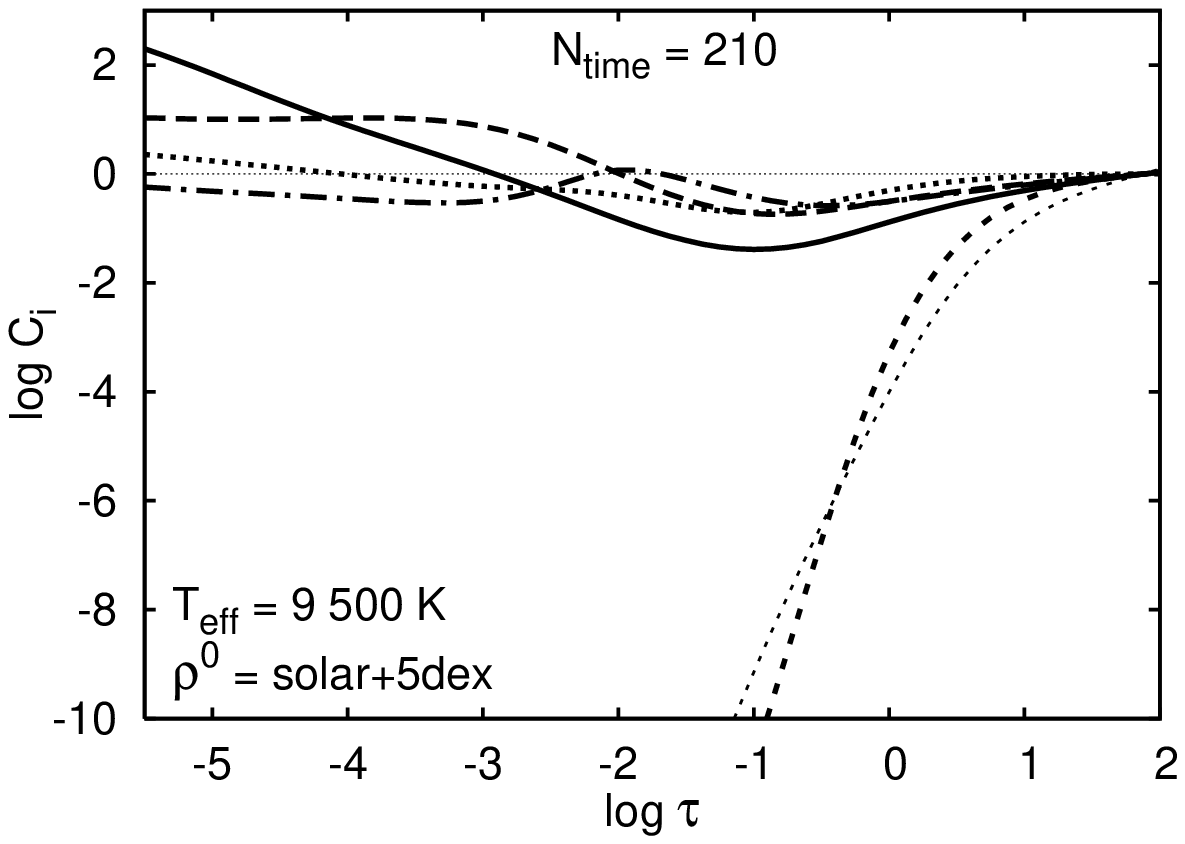}\includegraphics[viewport=5 15 355 252, clip, width=.5\textwidth]{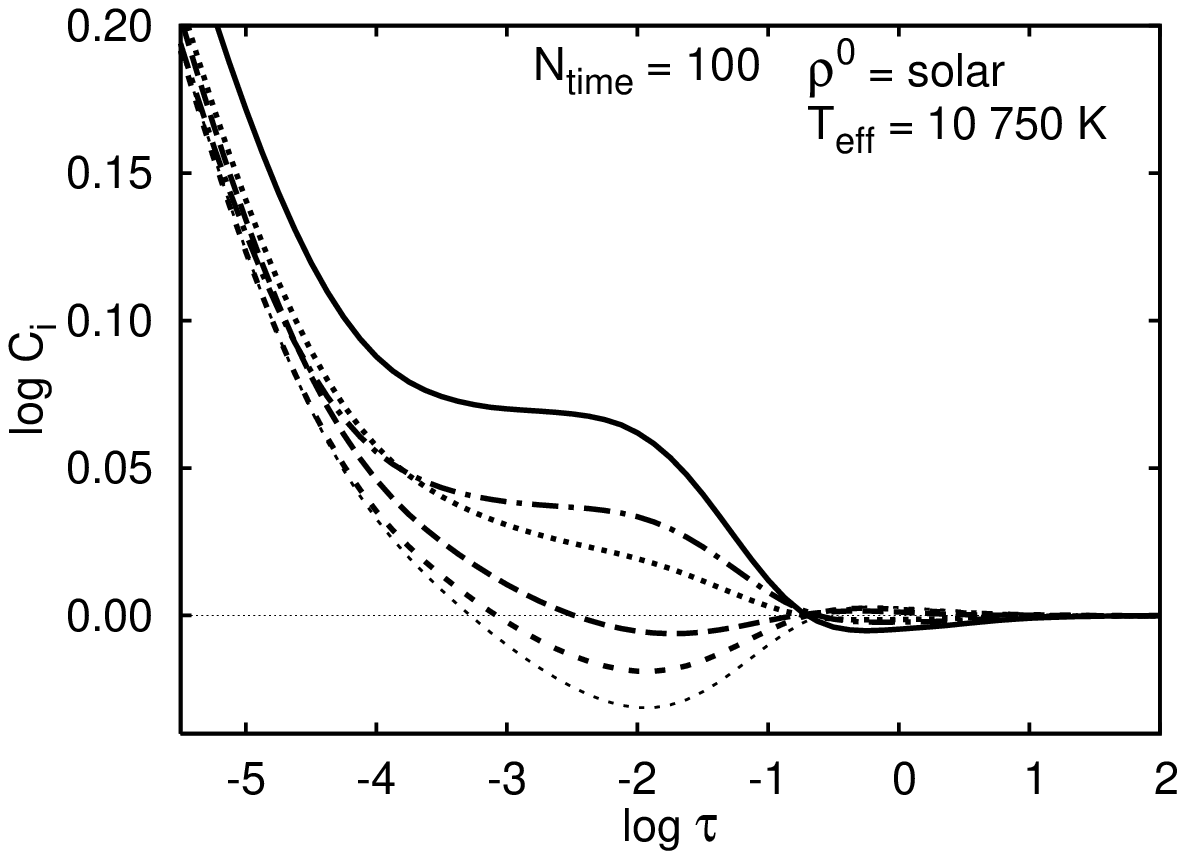} 
\includegraphics[viewport=5 15 355 252, clip, width=.5\textwidth]{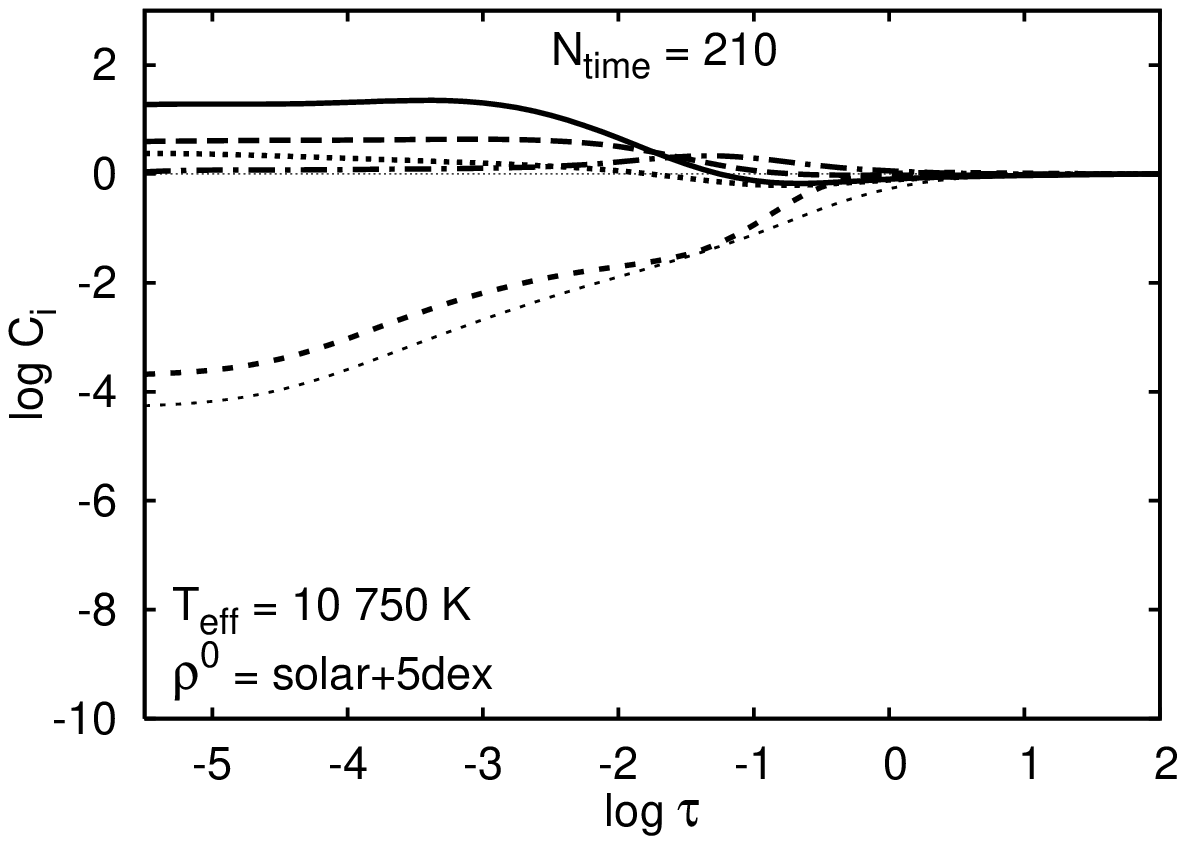}\includegraphics[viewport=5 15 355 252, clip, width=.5\textwidth]{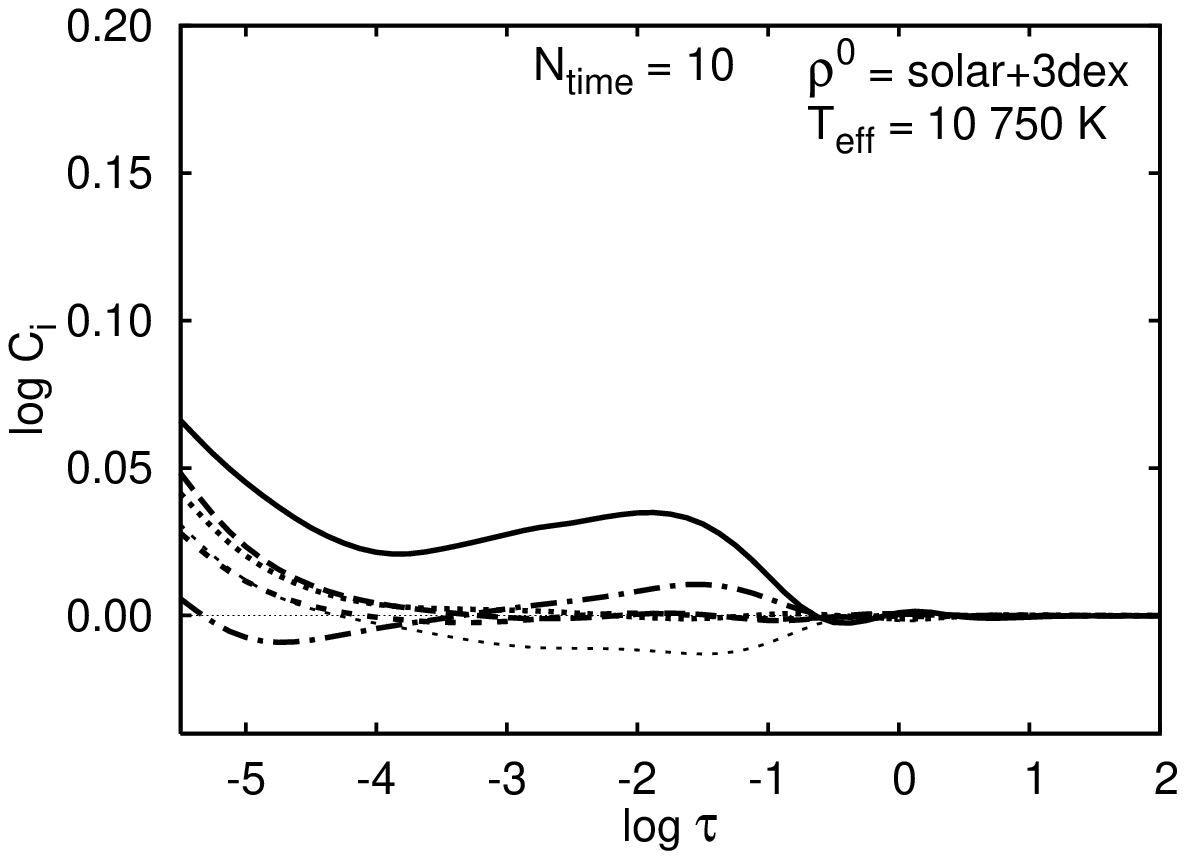} 
\includegraphics[viewport=5 -1 355 252, clip, width=.5\textwidth]{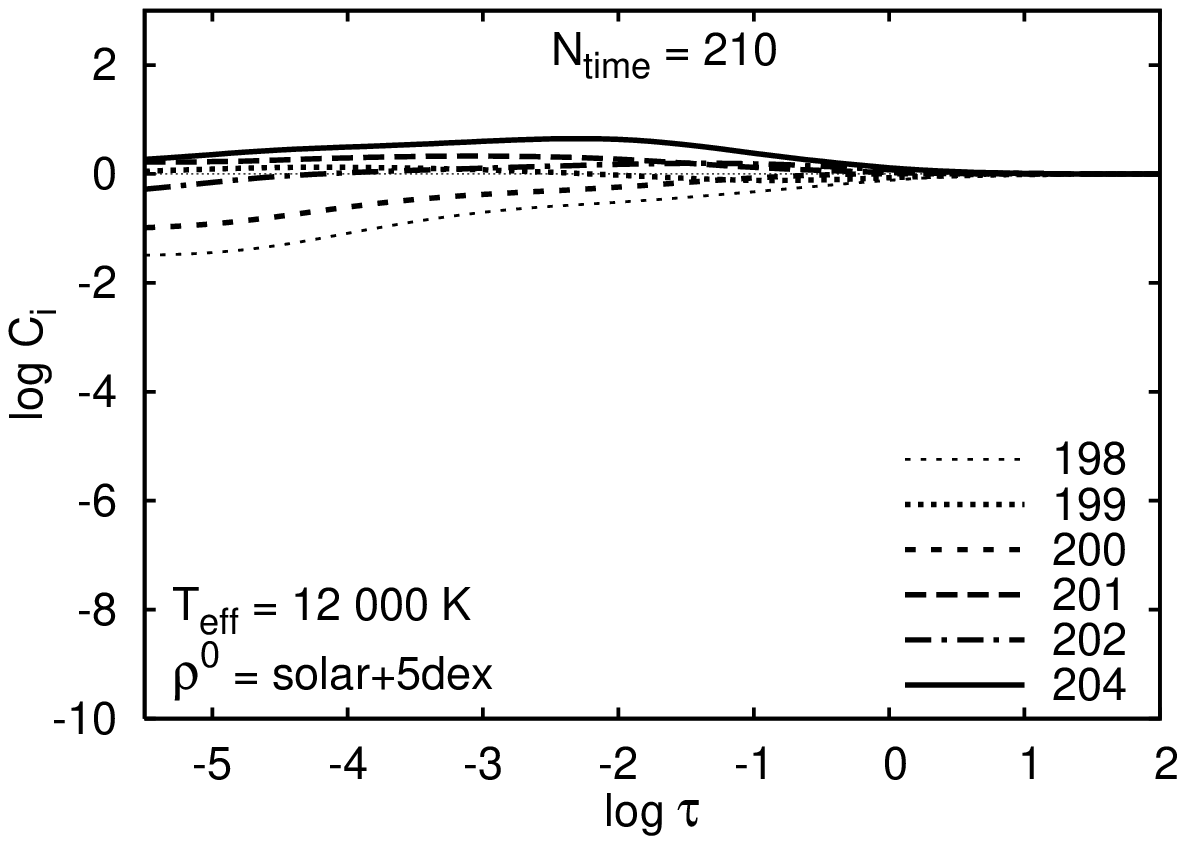}\includegraphics[viewport=5 -1 355 252, clip, width=.5\textwidth]{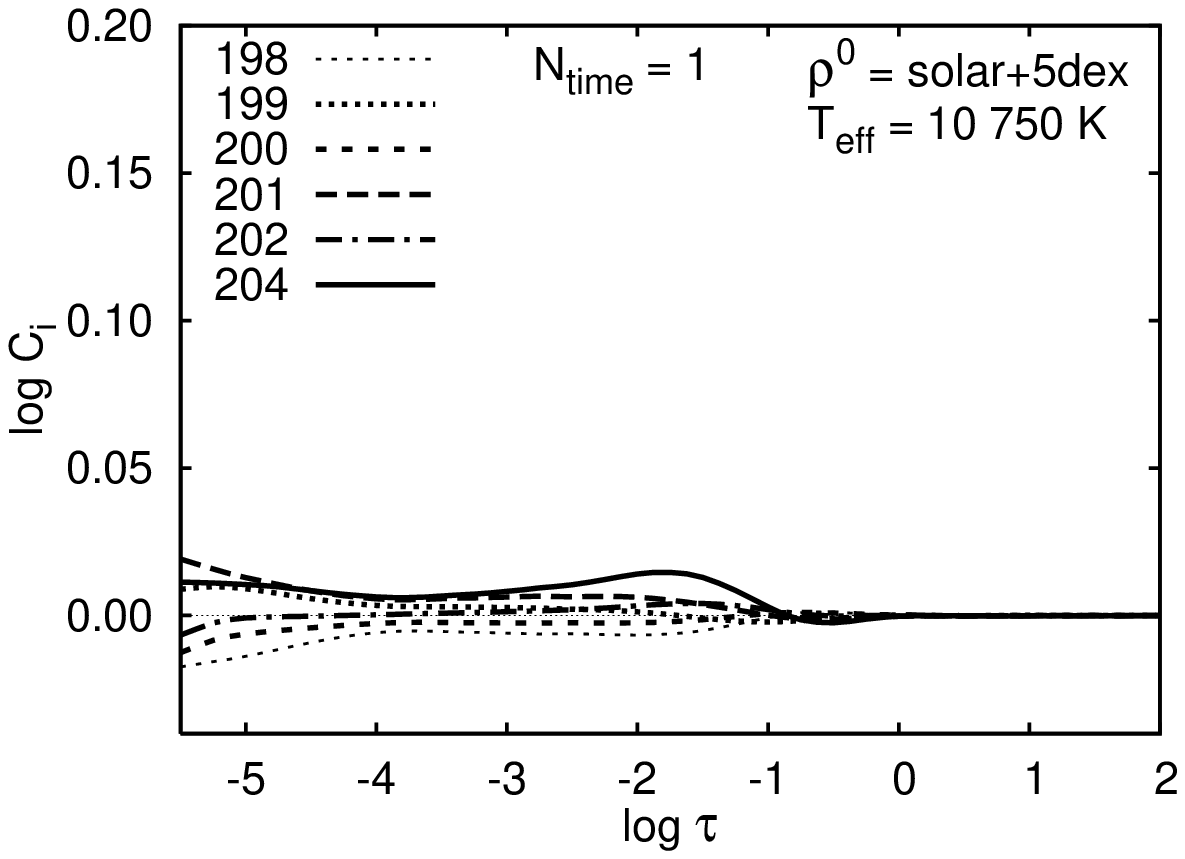}
  \caption{Evolutionary changes of Hg isotope concentrations relative
  to their initial values in logarithmic scale. Left column: $T_{eff}$ dependence (time=210 yr);
  right column: initial Hg abundance dependence (time=1 yr). Note the essential slowing of
  the segregation at higher $T_{eff}$ and Hg abundance values.}
  \label{fig:C_sapar2}
\end{figure}

Computed early evolutionary scenarios demonstrated rapid changes in the total
acceleration due to LID and presence of relaxational damping of isotopes flow.
The results are illustrated in Fig.\,\ref{fig:C_sapar1}, where a modified
logarithmic scale ${\rm sign}(a) \log\left(\left|\frac{a}{g}\right| +
1\right)$ has been  used for a sign-changing acceleration. The  dependence of
evolution of Hg concentration on  $T_{eff}$ has been illustrated in the left
column of Fig.\,\ref{fig:C_sapar2} and on the initial Hg abundance  -- in the
right column.

\begin{figure}[bth]
  \centering
\includegraphics[viewport=5 -1 355 252, clip, width=.5\textwidth]{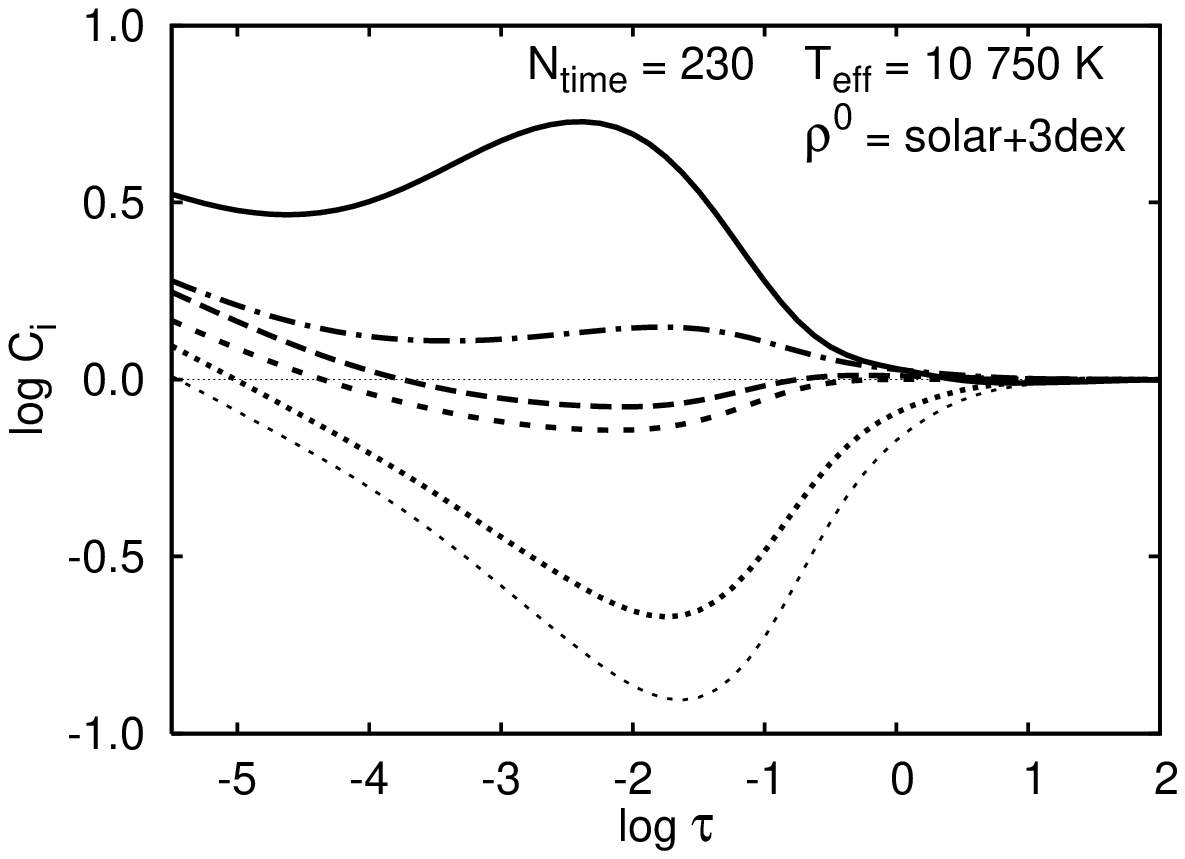}\includegraphics[viewport=5 -1 355 252, clip, width=.5\textwidth]{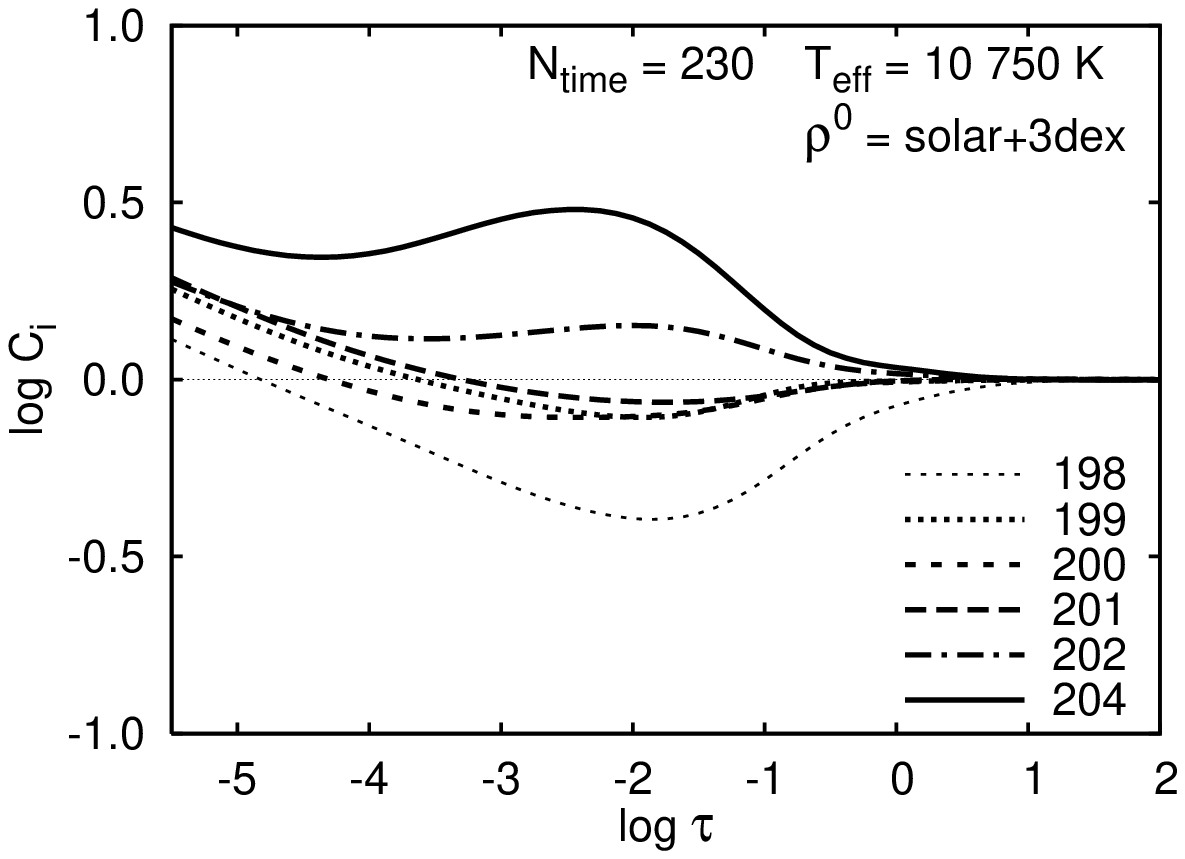}
  \caption{Evolutionary changes of "pure mercury" concentrations
    without hyperfine splitting (left panel) give the uncrossed curves
    in the order of isotope masses and with hyperfine splitting (right panel) give slower
    diffusion and  intersecting curves for intermediate isotopes.}
    \label{fig:C_sapar3}
\end{figure}

An evolutionary scenario has been computed also for the  "pure mercury" case,
where lines of all other elements were ignored. This evolutionary scenario has
been computed for model atmosphere with parameters $T_{eff}=10~750$ K, $ \log
g=4$ and initial Hg abundance  solar + 3dex, first ignoring  the hyperfine
splitting of spectral lines and thereafter taking it into account. The results
are illustrated in Fig.\,\ref{fig:C_sapar3}.

 \section{Main conclusions and future outlooks}
 The computed evolutionary sequences of isotope segregation can help to
 explain the observed unusual or ever enigmatic ratios of isotopes of heavy
 elements in the atmospheres of CP stars, including their vertical abundance
 profiles. Radiative acceleration is dominant at solar abundance of Hg, the
 role of LID increases with increase of Hg abundance and it becomes dominant
 throughout the atmosphere at Hg abundance about solar + 5dex. Separation of
 isotopes starts in the outer rarefied layers and thereafter extends into the
 deeper layers. The process proceeds much slower at higher $T_{eff}$ values
 and higher Hg abundances. Lighter isotopes with even number of nucleons sink
 rapidly. Hyperfine splitting of spectral lines of isotopes with an odd number
 of nucleons decelerates and weakens the  segregation of isotopes. It also
 causes mixing of the order of isotope spectral lines and thus makes the
 picture of evolutionary isotope segregation more complicated. The overlapping
 isotopic spectral line profiles are sensitive to isotope abundance throughout
 the whole atmosphere.

Several improvements are required to obtain more realistic evolutionary
scenarios. More complete and accurate data of spectral line strengths and
their damping constants, more exact cross--sections for  impact processes and
the physically more adequate initial and boundary conditions are needed.
Physically adequate inner boundary conditions can be specified only if deeper
layers of stellar envelope are included in modelling. More realistic time
scales can be obtained by including the processes which hinder the diffusion
(turbulence, stellar wind). Currently obtained diffusion time-scales
demonstrate only the maximal values for isotope segregation rates in
absolutely quiet stellar atmospheres.

There are several possibilities to continue the studies of evolutionary
segregation of isotopes of heavy elements. The simplest way is to compute
longer evolutionary sequences for mercury. There are also no essential
problems to include {stellar wind} into evolutionary computations (formulae
are given in Sapar {\it et~al.}, P27, these proceedings). A complicated
problem is to find physically correct  diffusion coefficient due to
microturbulence. Serious  problems are to elaborate new more stable
algorithms, enabling to integrate over longer time steps and to find the final
distribution of isotopes  without evolutionary computations.

\acknowledgements
We are grateful to Estonian Science Foundation for financial support by grant ETF 6105.

\end{document}